\documentclass[a4,amsmath, amssymb,prl,reprint]{revtex4-1}
\usepackage{graphicx}
\usepackage{color}
\usepackage{mathrsfs}
\graphicspath{{./figures/}}

\begin{document}
\title{Efficient Photonic Crystal Parametric Source harnessing high-Q resonances}
\author{Gabriel Marty$^{1,2}$ , Sylvain Combri\'{e}$^{1}$ , Fabrice Raineri$^{2,3}$ , Alfredo De Rossi$^{1}$}
\affiliation{$^{1}$ Thales Research and Technology, Campus Polytechnique, 1 avenue Augustin Fresnel, 91767 Palaiseau, France\\
$^{2}$Centre de Nanosciences et de Nanotetchnologies, CNRS, Universit\'{e} Paris Saclay, Palaiseau, France\\
$^{3}$Universit\'{e} Paris Diderot, Sorbonne Paris Cit\'{e}, 75205 Paris, France}

\begin{abstract}
A new tuning mechanism is introduced in high-Q multimode photonic crystal resonators allowing to harness the resonant enhancement of the parametric resonance systematically. As a consequence, ultra-efficient stimulated and spontaneous Four Wave Mixing at continuous microWatt pumping levels are observed, and the scaling with Q is demonstrated. Experimental results are in perfect agreement with an analytical model without fitting parameters.
\end{abstract}

\maketitle
\begin{figure}[h!]
 \includegraphics[width=1.0\columnwidth]{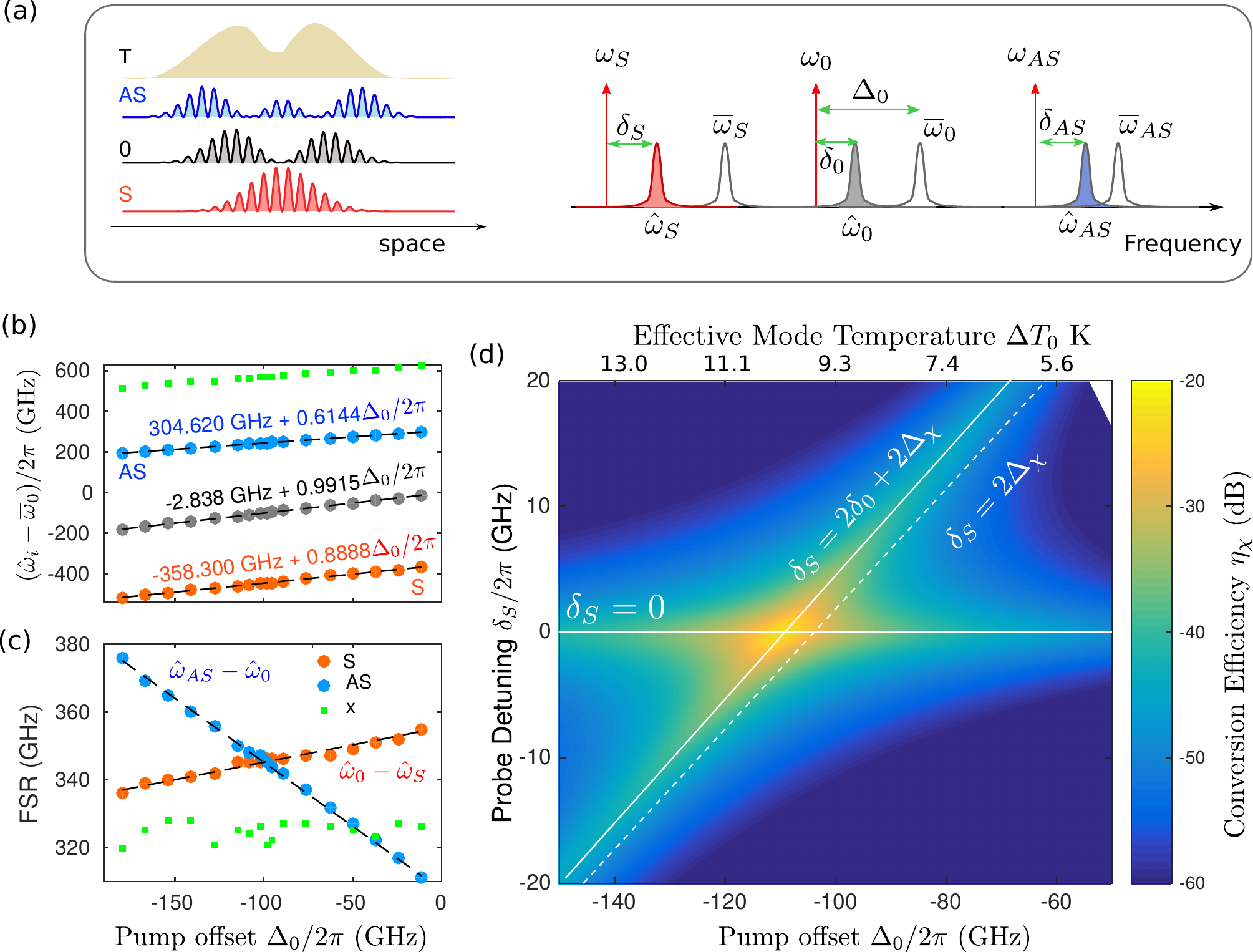}
\caption{\label{fig:frequency_matching}  (a) field $|E|^2$ of the eigenmodes and sketch of the thermal profile $T$ (left) and thermally shifted resonances relative to cold cavity eigenfrequencies and fields; (b) measured frequency of the first 4 eigenmodes (relative to $\overline\omega_0$) as a function of the pump offset $\Delta_0$ and corresponding linear fit; (c) corresponding eigenfrequency intervals; (d) calculated false color map of the efficiency of stimulated FWM $\eta_\chi$ as a function of the pump offset and probe detuning $\delta_S$ and effective temperature change $\Delta T_0$ for mode 0. The white  lines represent the poles of eq. \ref{eq:FWM_lowgain_alt}, and the  dashed is with $\delta_0=0$.}
\end{figure}

Spontaneous Parametric Down-conversion and Spontaneous Four Wave Mixing (SFWM), resulting from the nonlinear dielectric polarizability, underpin many nonclassical light sources. Their integration in photonic circuits is crucial for quantum optical technologies.\cite{caspani2017,harris2014} Silicon ring resonators, available from photonic foundries, have demonstrated high-quality time-energy entanglement with remarkable  brilliance \cite{grassani2015}, owing to the combination of small size and resonant enhancement. The miniaturization of the interacting modes could be pushed close to the diffraction limit \cite{laiGalli2014} in Photonic Crystal (PhC) cavities, resulting into a maximized Q-factor over mode volume ratio $Q/V$. This is a crucial ingredient for nonlinear optics at the single photon level \cite{Ferretti2012,volz2012}, parametric generation \cite{rivoire2009,galli2010,diziain2013,buckley2014,zeng2015,jiang2018} with ultra-efficient interaction at very low pump power levels \cite{linrodriguez2016}.\\
Energy conservation in resonant parametric interactions requires a strict control of the frequency spacing between interacting modes, which is challenging in PhC cavities\cite{Minkov2019doublyresonant,buckley2014nonlinear}. In the case of FWM, it requires at least three equi-spaced eigenfrequencies, a condition which must coexist with the diffraction-limited, low radiative losses design of the modes. Resonant FWM has been achieved  by combining three nano-beam PhC cavities with a moderately low Q factor($\approx 5\times10^3$) \cite{azzini2013stimulated}. Yet, harnessing the $Q^4$ scaling of the conversion efficiency in resonant Four Wave Mixing (FWM) \cite{helt2012} is very challenging, because a larger Q requires an increasingly stricter control of the eigenfrequencies. In fact, very large efficiency could be extrapolated considering the largest Q factor demonstrated in PhC ($Q\approx10^7$ \cite{asano2017}). However, without a proper tuning mechanism, they cannot be exploited as the structural disorder induces much larger statistical fluctuations of the eigenfrequencies (about 40 GHz \cite{taguchi2011}) than their spectral linewidth ($>20$ MHz).  Considering more than three coupled cavities \cite{matsuda2013,matsuda2017resonant} mitigates the impact of disorder, at a cost of a higher modal volume. Very recently, it has been shown that the photorefractive effect is very promising for tuning LiNbO$_3$ PhC cavities\cite{LiLin2019}.\\
Here we demonstrate a mechanism where three high-Q resonances of a photonic crystal cavity are tuned dynamically to form an equally spaced triplet. The resulting frequency alignment allows to reach unprecedented ultra efficient parametric interaction in this type of resonators. Interestingly, the other cavity modes remain misaligned and, therefore, they do not interact. Consequently, the system can be fully described by a simple analytical model of an ideal degenerate parametric system. It was suggested that such a system could be used as a ``noise eater''\cite{matsko2019}. Our system allows the selective activation of specific parametric interactions, which has been achieved in corrugated ring resonators\cite{luLin2014} and it is very important in the context of quantum communications\cite{luSrinivasan2019}. With a loaded Q factor $>1\times10^5$ and a mode volume in the $(\lambda/n)^3$ range, we reach the predicted conversion efficiency (-6 dB) with on-chip continuous pump power below 100$\mu W$. Under the same conditions, SFWM is bright enough to be easily measured with an Optical Spectrum Analyzer. Its power level is only 60 dB below the pump, which greatly eases on chip filtering compared to ring resonators. All experimental data are in excellent agreement with the prediction of our model and the scaling of the parametric interaction with the Q factor is demonstrated.\\
\newline
%
%
\begin{figure}[h!]
 \includegraphics[width=1.0\columnwidth]{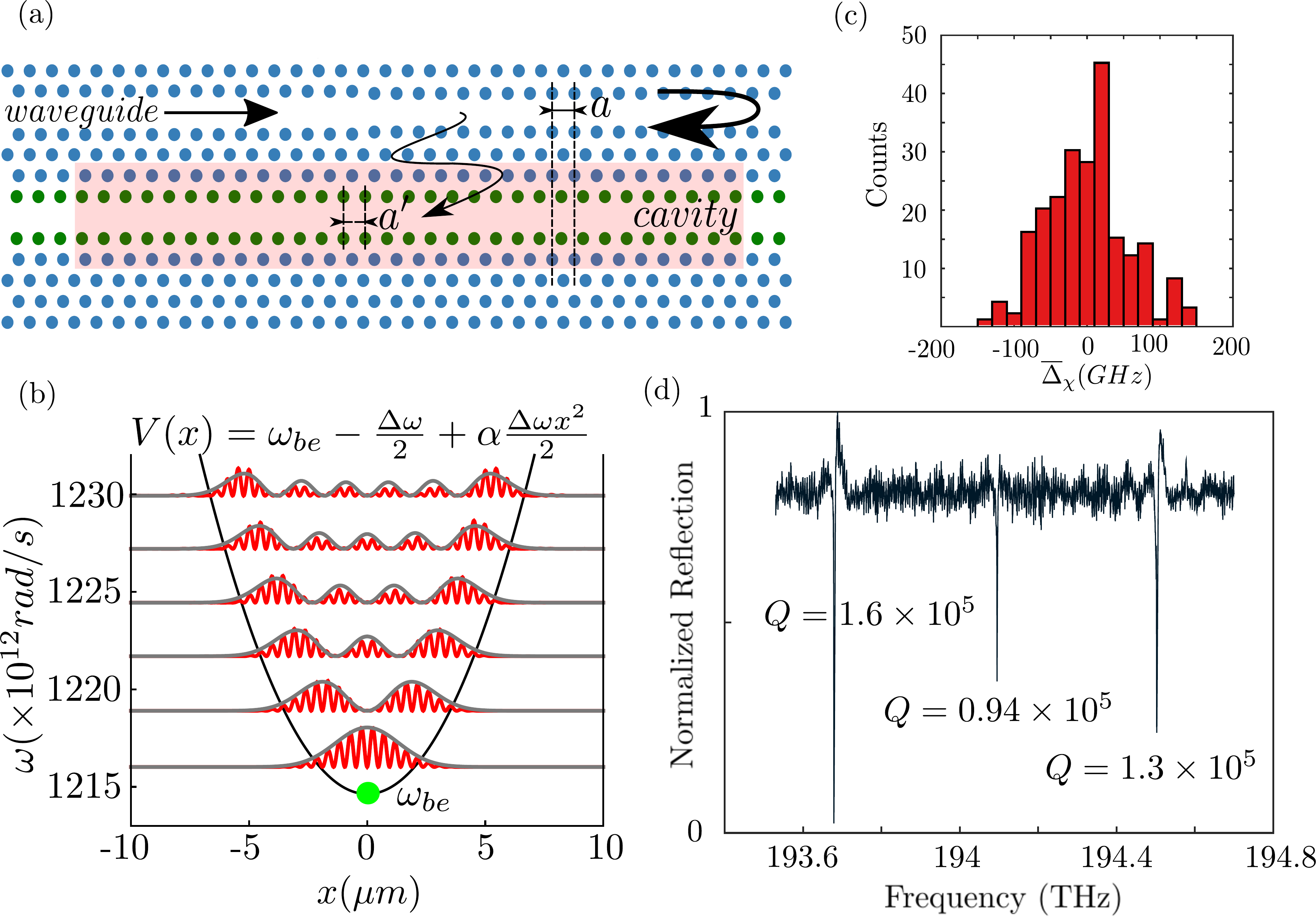}
\caption{\label{fig:optical cavity}  (a) Geometry of the bichromatic Photonic Crystal Cavity; (b) Mode envelopes fitted with Hermite Gauss functions(c)Measured histogram of eigenfrequencies misalignment ($\overline\Delta_\chi/\pi$) (d) Measured Linear scattering spectrum of cavity with original misalignment $\overline\Delta_{\chi}$ = - 28 GHz  and $Q_{avg} = 128 000$.}
\end{figure}
%
%
%
%
The optical cavity consists in a thin slab of InGaP patterned with a bichromatic lattice, Fig. \ref{fig:optical cavity}(a). This pattern creates an effective photonic potential, Fig. \ref{fig:optical cavity}(b) that maximizes the $Q/V$ ratio of the optical mode\cite{alpeggiani_effective_2015,simbula2017}. As shown in Ref. \cite{combrie2017}, it corresponds to an effective harmonic potential as the cavity eigenmodes are equispaced in frequency with envelopes described by the Hermite-Gauss functions. Furthermore, the Maxwell Equations can be approximated by a Schr{\"o}dinger equation\cite{SipeWinful1988} near the band edge ($\omega_{be}$) of the photonic band, namely:
\begin{equation}
\imath\partial_t A -\frac{1}{2}\omega_{kk} \partial_x^2A +V(x) A - \alpha |A|^2A=0
\label{eq:GPE}
\end{equation}
Here  $\omega_{be}+\omega_{kk}k^2$ represents the dispersion of the photonic conduction band, $A$ is the envelope of the Bloch modes, $V(x)$ is the effective potential and $\alpha$ is a third order nonlinear response\footnote{Y. Sun, S. Combri\'e, F. Bretenaker and A. de Rossi, submitted}.\\
In contrast to whispering gallery modes in toroids and spheres, or propagating waves in ring resonators, the spatial distribution of the energy density in optical cavities with stationary modes, such as PhC resonators, is not uniform. Each mode have a specific spatial pattern (see Fig. \ref{fig:optical cavity}b). As a consequence, structural disorder (i.e. any deviation from the design such as roughness or other fabrication imperfections) induces almost uncorrelated  fluctuations in the eigenfrequencies. 
Therefore, fluctuations on the spectral alignment three of consecutive resonances $2\overline\Delta_\chi=2\overline\omega_n-\overline\omega_{n+1}-\overline\omega_{n-1}$ are on the same order of magnitude. Statistics on 96 cavities (260 triplets) shows that the standard deviation of $\overline\Delta_\chi/\pi$ can be estimated to be about 50 GHz, Fig. \ref{fig:optical cavity}(d), which is orders of magnitude larger than the minimum linewidth achievable in PhC cavities (about 20 MHz)\cite{asano2017}.\\
%
Consequently, it is absolutely necessary to find a mechanism to compensate $\overline\Delta_\chi$, which implies the control of multiple narrow resonances on a span which is several orders of magnitude larger than their linewidth. We propose a solution that exploits the very same peculiarity of our system that is responsible for large $\overline\Delta_\chi$, namely the different spatial structure of the eigenmodes. A non uniform perturbation of the refractive index will in fact affect the eigenmodes differently, resulting into a change of $\overline\Delta_\chi$. Independent cavity mode tuning of multiple coupled PhC cavities has been achieved through local thermorefractive effect, where a well defined temperature profile is generated with a holographic laser pattern\cite{SokolovLian2017,YuceLian2018}. Here this effect is exploited in order to manipulate the modes within a single cavity.\\ 
Let us now consider the first three eigenfrequencies of our cavity with a frequency mismatch $\overline\Delta_\chi>0$, with spatial profiles sketched in Fig. \ref{fig:frequency_matching}(a). Let us label the modes in order of increasing frequency as ``S'' (Stokes), ``0'' and ``AS'' (Anti-Stokes) and suppose mode ``0'' is resonantly excited with a laser, initially at frequency $\omega=\overline\omega$. When the laser is red-detuned, the eigenfrequency will also red-shift  because it is thermally locked \cite{Carmon2004} and its effective temperature\cite{sokolov2015} will increase by $\Delta T_0\approx\Delta_0/\partial_T\omega$ with the laser frequency offset $\Delta_0=\omega_0-\overline\omega\approx\hat\omega_0-\overline\omega_0$.
It is noted that in a ring resonator, or in any traveling wave resonator, the spectral shifts of the eigenfrequencies are highly correlated, because they are all subject to the same temperature change. With Hermite-Gauss modes, the temperature profile $T(\mathbf{x})$ is inhomogeneous because it follows the energy distribution $|\mathbf{E}(\mathbf{x})|^2$ of the excited mode (here mode ``0''). Two local hot spots are created in the cavity. More precisely, $T(\mathbf{x})$ is spatially broadened with a diffusion length $\approx 4$ $\mu m$.\cite{sokolov2015}, small enough to affect the eigenmodes differently (Fig. \ref{fig:optical cavity}). As a consequence, modes ``S'' and ``AS'' will experience a different spectral shift as the mode ``0'' is dragged by the pump. 
Using an optical homodyne technique \cite{combrie2017}, the eigenfrequencies are measured while mode ``0'' is detuned by the laser with on-chip power about 800 $\mu$W, as shown in Fig. \ref{fig:frequency_matching}(b). The detuning of each mode is linear with respect to $\Delta_0$, but their slopes are different, as expected. Since each resonance frequency is changed by a different thermo-optic coefficient, there is a value of $\Delta_0$ which aligns the triplet ``S'', ``0'', ``AS'', namely $2\Delta_\chi=2\hat\omega_0-\hat\omega_S-\hat\omega_{AS} =0$, $\hat\omega$ denoting the detuned eigenfrequency. Interestingly, this condition is not satisfied simultaneously for the next higher mode (labeled ``X''), as shown in Fig. \ref{fig:frequency_matching}(c). This is a peculiarity of this tuning mechanism. It can also be noted that there is no detectable increase of the linewidth as the laser is swept, implying that absorption is negligible compared to scattering losses.\\
%
%
%
%
%
Based on the dependence of the Free Spectral Range (FSR) on the pump offset $\Delta_0$ measured here, the theoretical efficiency $\eta_\chi=P_{id}^{out}/P_{s}^{in}$ (idler over signal power) of the stimulated FWM process is predicted as a function of the pump frequency (to which mode ``0'' is locked) and probe detuning $\delta_S=\omega_S-\hat\omega_S$ from mode ``S''. This is represented by the false-color map in Fig. \ref{fig:frequency_matching}(d), which is calculated by solving the master equation in the undepleted pump approximation 
for a triply resonant cavity. The efficiency $\eta_\chi$ reaches a maximum when $\delta_{S}=\delta_{0}=\delta_{AS}=0$, which is only possible if $\Delta_\chi=0$. The maximum is given by:
\begin{equation}
\eta_\chi^{(max)}=\left[\frac{c_0 n_2 \omega}{\varepsilon_r V_{\chi}}\right]^2  \frac{4^4\kappa_{0}^2}{\Gamma_{0}^4}\frac{\kappa_{S}\kappa_{AS}} {\Gamma_{S}^2\Gamma_{AS} ^2}P_{0}^2
\label{eq:STPC_max}
\end{equation}
Here $n_2=0.6\times10^{-17}m^2W^{-1}$ is the nonlinear index of the material, $V_\chi=27 V_m$ corresponds to the calculated field overlap of the three interacting modes, $V_m\approx 2(\lambda/n)^3$ is the volume of the fundamental mode, $\kappa$ represents the coupling rate to the waveguide and $\Gamma$ the total photon decay rate. $P_0$ is the on-chip power reaching the waveguide.
When the detuning is not zero and in the limit $\eta_\chi\ll 1$ the efficiency can be rewritten as:
\begin{align}
\eta_\chi=\eta_\chi^{(max)}\mathcal{L}\left(\frac{\delta_{0}}{\Gamma_{0}}\right)^2\mathcal{L}\left(\frac{\delta_{S}}{\Gamma_{S}}\right)\mathcal{L}\left(\frac{2\delta_{0}+2\Delta_\chi-\delta_{S}}{\Gamma_{AS}}\right)
\label{eq:FWM_lowgain_alt}
\end{align}
having defined the Lorentzian $\mathcal{L}(x)=1/(1+4x^2)$.  
$\Delta_\chi$ is extracted from Fig. \ref{fig:frequency_matching}(c).

Considering that the pump mode ``0'' is quasi-resonant, this equation admits two possible local maxima, when either $\delta_S=0$ or $\delta_S=2\delta_0+2\Delta_\chi$. The absolute maximum is reached when both conditions are satisfied, meaning that the cavity is spectrally aligned and the probe is resonant.  
\\
%
%
\begin{figure}[h!]
 \includegraphics[width=1.0\columnwidth]{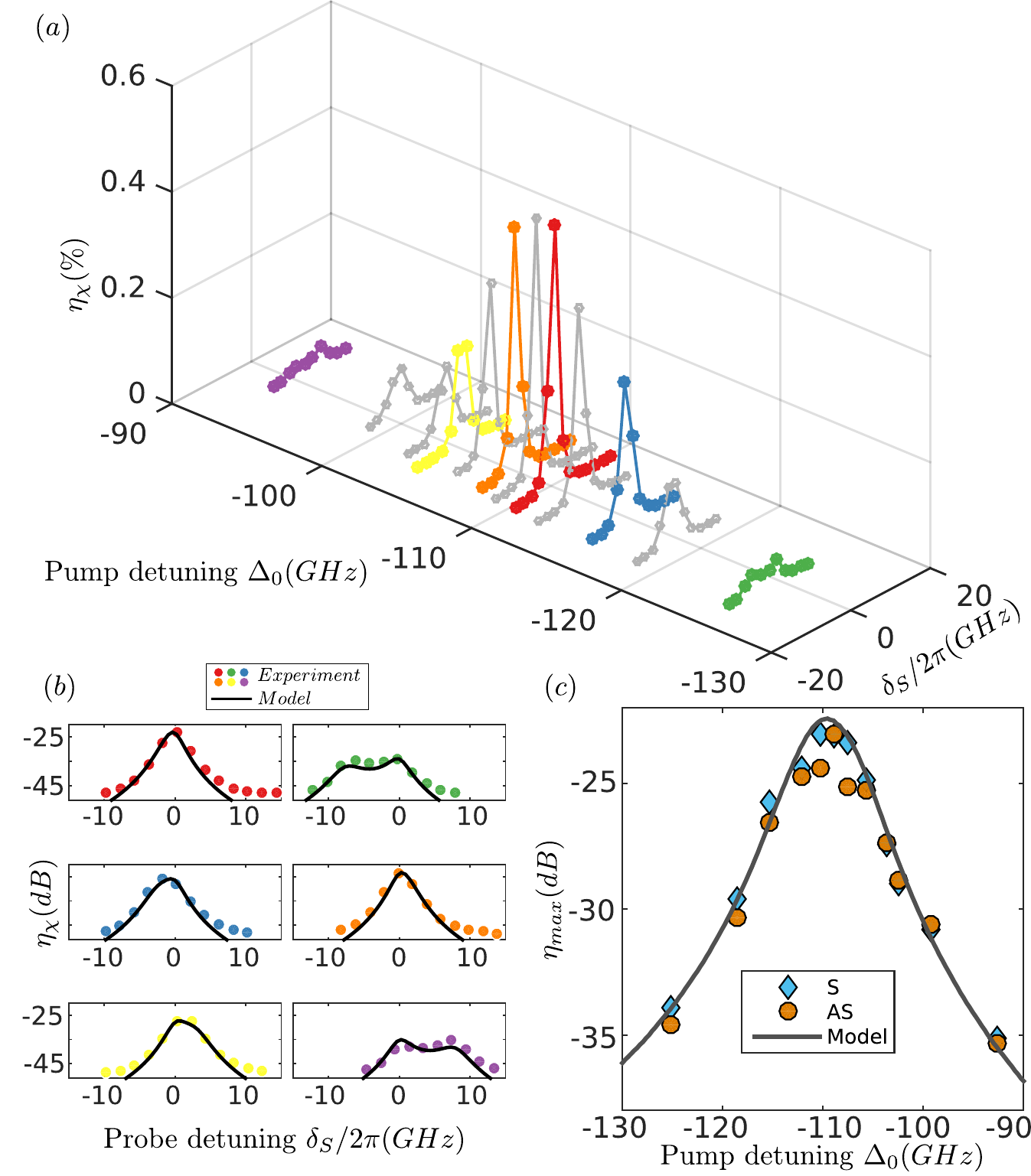}
\caption{\label{fig:FWM} Measurement of the stimulated FWM efficiency $\eta_\chi$ as a function of the pump offset $\Delta_0$ and probe detuning $\delta_S$; the corresponding measured FSR is in the inset (a); comparison with the model (inset with colored frame); (b) $\eta_{max}$ as a function of the pump offset, experiment (symbols) and theory (line).}
\end{figure}
Stimulated FWM measurements, shown in Fig. \ref{fig:FWM}, have been performed with two narrow linewidth ($<0.1 MHz$), mode-hopping free tunable semiconductor lasers; the signal laser sweeps either the S or the AS eigenmode while keeping the pump detuning fixed. The procedure is repeated for a different pump detuning but keeping the on-chip pump level at 700 $\mu W$. The signal power level is sufficiently low that it does not induce additional thermal bistability. The conversion efficiency $\eta_\chi$ is extracted from the raw Optical Spectrum Analyser (OSA) traces recorded for each combination of pump and signal settings. The cavity is coupled with a single-end waveguide (\ref{fig:optical cavity}a) that reflects all the signals, and a circulator is used to filter the input and the output.  
The reflected idler power represents  $P_{id}^{out}$ and the reflected signal, when tuned out of resonance, represents $P_{s}^{in}$. This is justified because nearly total reflection occurs on when the input is out of resonances and the propagation losses in the short coupling waveguide are negligible.\\
As shown in Fig. \ref{fig:FWM}(a),a maximum efficiency is obtained at the offset $\Delta_0\approx-115$ GHz, when FSR are equal to 345 GHz, as extracted from the same OSA spectra (inset). It corresponds approximatively to the FSR crossing in Fig. \ref{fig:frequency_matching}(c). As shown in Fig. \ref{fig:frequency_matching}(d), the maximum is red shifted because of the pump is here slightly detuned from the hot resonance ($\delta_0\approx1 GHz$). This confirms that the interacting modes have been tuned to maximize the parametric interaction. Two local maxima, gradually separating each other appear as $\Delta_0$ is further increased, consistently with Fig.~\ref{fig:frequency_matching}(d). This is more apparent in Fig. \ref{fig:FWM}(b), considering the dependence of $\eta_\chi$ on the probe detuning $\delta_S$ (similar is obtained when tuning around ``AS''). The quantitative agreement with theory is very good. We note that all the parameters used in the model have been measured independently or calculated, except for the thermo optic coefficient $\alpha_{00}$ and the pump power which have adjusted slightly \footnote{The measured value is $700 \mu W$ and the value used in the model is $800\mu W$}.
The dependence of $\eta_{max}=\max[\eta_\chi(\delta_S)]$ is shown in Fig. \ref{fig:FWM}(c) 
as a function of $\Delta_0$. As expected, the measured $\eta_{max}(\Delta_0)$ is the same when tuning the probe either on the ``S'' and on the ``AS'' eigenfrequency, as expected; the agreement with theory is also very good.\\
The properties of non-classical light generated through the spontaneous FWM process can be characterized indirectly from the measurement of the stimulated FWM\cite{Liscidini2013}. In particular, the power level of the spontaneous emission has been calculated for ring resonators and shown to be related to the classical formula for the stimulated FWM\cite{helt2012}. Consistently with Ref. \cite{chembo2016}, the link between spontaneous emission from mode $m$ and stimulated conversion efficiency $\eta_{\chi}(\Omega)$ can be expressed in the general case as 
\begin{equation} 
P_{SP,m}=\frac{\hbar\omega_m}{2\pi}  \frac{\kappa_m}{\Gamma_m}\int{\eta_{\chi}(\Omega)}d\Omega
\label{eq:Pspe}
\end{equation}
where $\Omega=\omega-\hat\omega$ denotes the spectral detuning from the Stokes (or anti-Stokes) resonance. It has been shown that $P_{SP}$ is proportional to $Q^3/V_{\chi}^2$, which is exactly the same scaling as in ref \cite{helt2012}. The term $\kappa_m/\Gamma_m$ represents the fraction of internally generated pairs which are not lost due to internal losses and implies that the emission rate from each mode is in general different.\\
\begin{figure}
 \includegraphics[width=1.0\columnwidth]{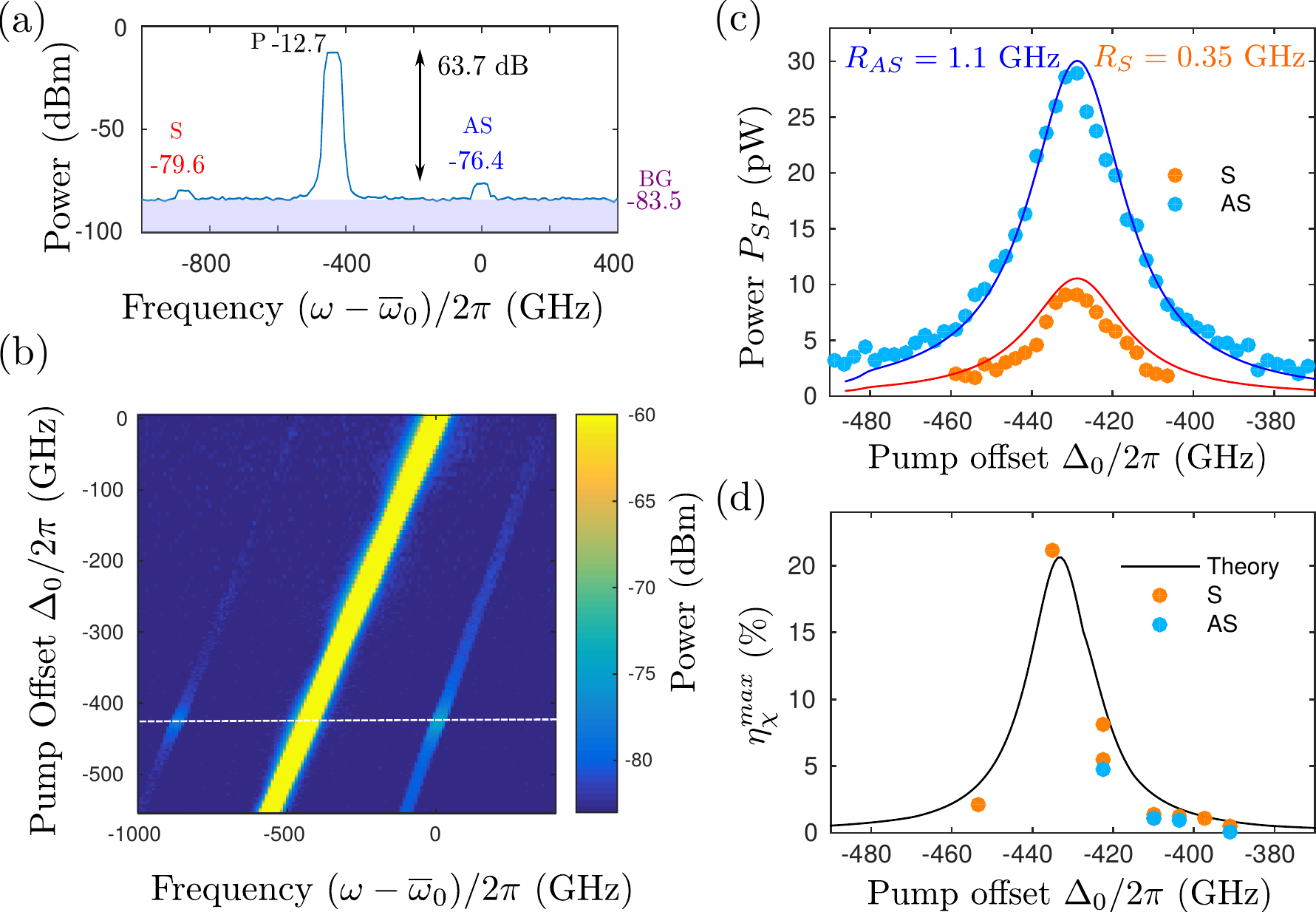}
\caption{\label{fig:FWM_spont} Sample with larger Q and $\overline\Delta_\chi$. (a) raw spectrum of the spontaneous emission (S, AS) the pump (P); the noise floor is represented by the shaded area, the levels are indicated. (b) Raw spectra (resolution is 100 GHz) as a function of the pump offset represented by a false-color map. The dashed line corresonds to the plot in (a). (c) Spontaneous FWM on the Stokes and anti-Stokes side as a function of the pump offset, theory accounts for 7dB insertion loss. (d) Maximum $\eta_\chi$ as probe is tuned on the Stokes or anti-Stokes vs. pump offset, theory (solid line) and measurements ( symbols).}
\end{figure}
It will be shown that the spontaneous emission rate is large enough to be measured with an OSA : as the pump level is here below 1 mW, the dynamic range of the OSA is sufficient ot measure the pump and the SFWM simultaneously. Thus spontaneous and stimulated FWM can be compared with a very simple procedure. For this measurement, a resonator with slightly larger Q factor 
 and larger eigenfrequency mismatch $\overline\Delta_\chi$ has been considered. Linear measurements show that mode ``0'' needs to be pulled by 430 GHz until $\Delta_\chi=0$ with on-chip power level of 700 $\mu W$. 
Fig. \ref{fig:FWM_spont}a,b represent the raw OSA traces recorded when the probe is switched off. As the pump offset is adjusted close to -430 GHz, the spontaneous emission is easily measured from the ``S'' and ``AS'' modes. We note that no optical band-pass filter is used here. The power level of the spontaneous emission is extracted from the raw measurements and plot as a function of the pump offset$\Delta_0$ in Fig. \ref{fig:FWM_spont}c. The agreement with the theory (eq. \ref{eq:Pspe} and 
experiment is very good. The maximum spontaneous FWM power measured at the OSA from the ``S'' mode is about 30 pW (150 pW on chip), corresponding to an on-chip emission rate of about 1 GHz. Remarkably,  spontaneous FWM is responsible for the high background in Fig. \ref{fig:FWM_spont}(a,b) when the probe is detuned from resonance, explaining the discrepancy with the calculated $\eta_\chi$. Moreover, the pump power is only 60 dB higher than the generated pair power owing to the very high efficiency of the parametric process. This is of particular interest when on-chip filtering is needed. Eq. \ref{eq:Pspe} suggests a relationship with the stimulated emission which could be approximated as: $P_{SP,m}=\frac{\hbar\omega_m}{2\pi}  \frac{\kappa_m}{\Gamma_m}\eta_{\chi}^{max}\Gamma_{sp}$, where $\eta_{\chi}^{max}=max[\eta_{\chi}(\Omega)]$ and $\Gamma_{sp}$ is a generation bandwidth\cite{helt2012} for the spontaneous process. This is apparent when comparing with the measurement of the corresponding simulated FWM efficiency, Fig. \ref{fig:FWM_spont}(d). Again, the agreement with the theoretical prediction is very good, also noting that the parameters have been measured independently measured or calculated 
and the pump power adjusted to $760 \mu W$. The stimumated conversion efficiency reaches $20\%$ (-7dB).\\
%
\begin{figure}
 \includegraphics[width=0.95\columnwidth]{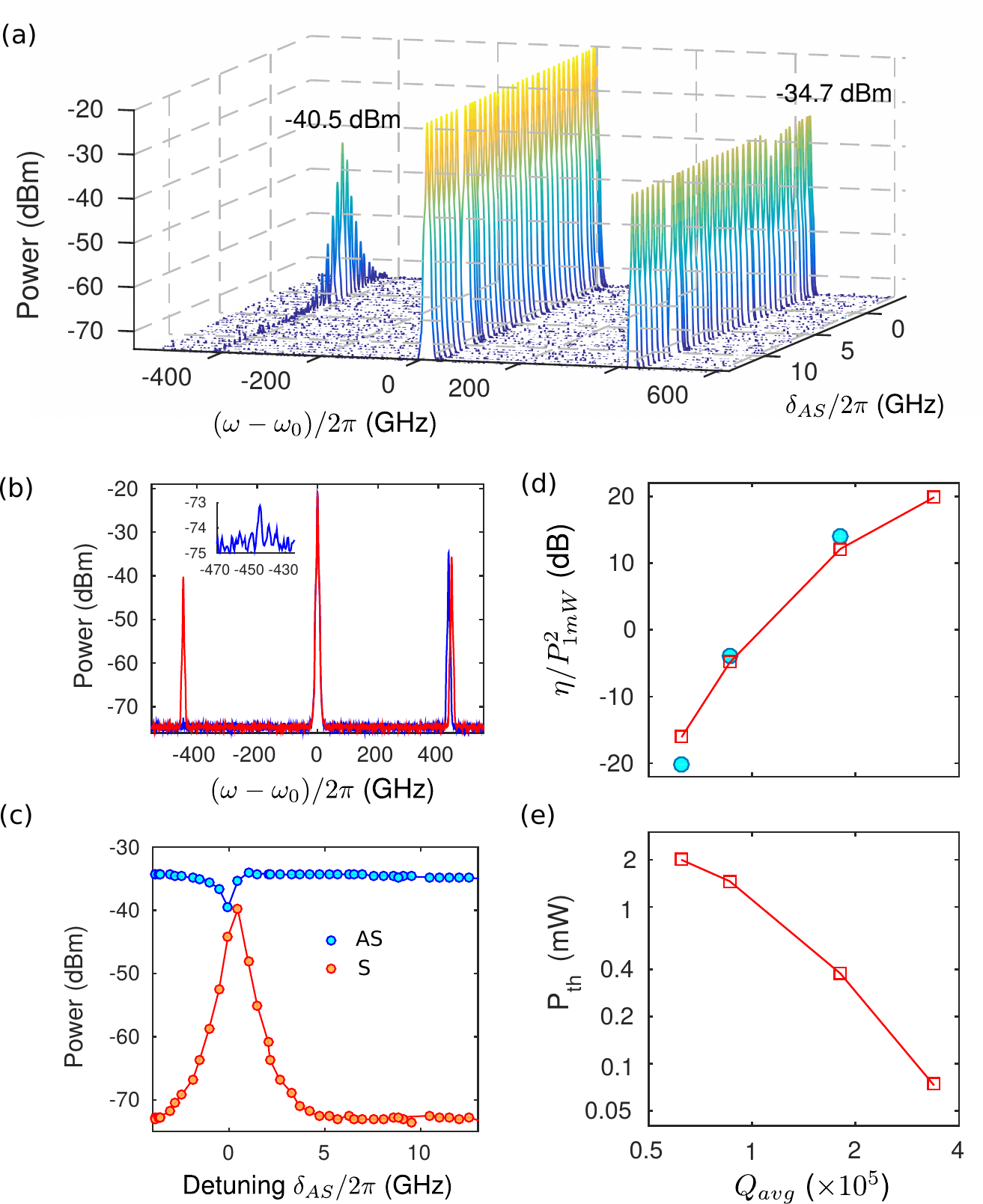}
\caption{\label{fig:FWM_scaling} Resonator with larger $Q_{avg}$. Raw spectra centered on the pump $\omega_0$ as a function of the probe detuning $\delta_{AS}$ (a); spectra for tuned (red) and detuned (blue) probe (b), dependence of the signal and idler on the detuning (c); calculated (solid line and squares) and measured normalized FWM efficiency (filled circles)as a function of $Q_{avg}$ for the three resonators considered here and a fourth with even larger $Q_{avg}$ (d); corresponding predicted threshold for optical parametric oscillation (e).}
\end{figure}
The very large efficiency normalized to the pump power level of FWM interaction results from the combination of large material nonlinear susceptibility, small interaction volume and the capability to exploit the strong scaling with the Q factor. In order to further illustrate this point, stimulated FWM has been measured in a third cavity with even larger Q. Here $\eta_\chi= -5.8dB$ is achieved at an on chip pump level of 80 $\mu W$. The corresponding raw spectra are shown in Fig. \ref{fig:FWM_scaling}(a) as a function of the probe detuning $\delta_{AS}$ from the Anti-Stokes resonance. The spectra at maximum $\eta$ is shown in Fig. \ref{fig:FWM_scaling}(b) along with the case where the probe is detuned far off resonance. In this case SFWM peak appears (inset). The ``S'' (idler) and ``AS'' peaks are traced as a function of the detuning, Fig. \ref{fig:FWM_scaling}(c). The very large FWM efficiency, normalized to the pump power at 1 mW, namely $\eta/P_{1mW}^2$ is explained in terms of the larger $Q_{avg}=(Q_S Q_{AS} Q_0^2)^{1/4}$ (see Table III in Supplementary). This follows very well the scaling in eq. \ref{eq:STPC_max}, which is plot in Fig.~\ref{fig:FWM_scaling}(d) along with experimental points (the three resonators considered here) and the prediction for with even larger Q measured in another cavity. Finally, the corresponding threshold of optical parametric oscillation predicted for these cavities is plot in Fig.~\ref{fig:FWM_scaling}(e) and reveals a very strong decrease of the threshold pump power for the onset of optical parametric oscillations.\\  
\newline
In conclusion, we have demonstrated that a photonic crystal cavity with mode volume $V_m\approx 2(\lambda/n)^3$ can be controlled to form a triplet of equi-spaced resonances. Because of the strong inhomogeneity of their spatial distribution, the thermo-refractive effect shifts their resonances differently. Therefore, a combination of specific pump power and detuning can achieve spectral alignment, even if the modes are initially strongly misaligned because of fabrication imperfections. In these conditions, stimulated FWM becomes extremely efficient, following the predicted scaling with $Q^4$. Thus, a maximum $\eta_{FWM}=-5.8dB$ is achieved with a pump level of $80 \mu W$, as predicted for the measured  Q factor $2\times10^5$. Spontaneous emission is also very strong, with GHz emission rates that already compares with state of the art silicon nitride ring resonator, with a lower pump to filter. We have compared our measurements with an unified model where only the coupled pump level has been allowed to be adjusted slightly, relatively to the measurements. An excellent agreement is obtained systematically, demonstrating that the variety of measurement performed on different resonators are all consistent with the expected scaling of the interaction with the Q factor. This implies that FWM could harness the record high Q ($\approx10^7$) demonstrated in PhC resonators, leading to efficient interactions and parametric oscillation at few $\mu W$ power levels. Finally, the tuning mechanism demonstrated here will allow only a triplet of modes to interact, implementing therefore an ideal parametric system where, for instance, ultra low noise signal processing and quantum manipulation could be performed.

\bibliography{../biblio/parametric_sources.bib,../biblio/quantum_sources.bib,../biblio/parametric.bib,../biblio/photonic_crystals.bib,../biblio/PBG_TRT.bib}


\end{document}